\documentclass[12pt,a4paper]{article}
\usepackage[english]{babel}
\usepackage{latexsym,amssymb,amsmath}
\usepackage{graphicx}
\usepackage{geometry}
\usepackage{mathptmx}
\usepackage{wrapfig}

\begin{document}

\title{ Effect of the heliospheric interface on the distribution
of interstellar hydrogen atom inside the heliosphere}
\author{
$\mbox{O.A. Katushkina}^{*}$ (1,2), V.V. Izmodenov (1,2,3) \\
(1) IKI RAS, 117997, Moscow, Profsoyuznaya Str 84/32 \\
(2) Lomonosov Moscow State University, 119992, Moscow, GSP-1, Leninskie Gory \\
(3) IPMech RAS, Moscow, prosp. Vernadskogo, 101-1
}
\date{}

\maketitle
\begin{abstract}
    This paper deals with the modeling of the interstellar hydrogen atoms (H atoms) distribution in the heliosphere. We study influence of the heliospheric interface, that is the region of the interaction between solar wind and local interstellar medium, on the distribution of the hydrogen atoms in vicinity of the Sun. The distribution of H atoms obtained in the frame of the self-consistent kinetic-gasdynamic model of the heliospheric interface is compared with a simplified model which assumes Maxwellian distribution of H atoms at the termination shock and is called often as 'hot' model. This comparison shows that the distribution of H atoms is significantly affected by the heliospheric interface not only at large heliocentric distances, but also in vicinity of the Sun at $\sim 1-5$ AU. Hence, for analysis of experimental data connected with direct or undirect measurements of the interstellar atoms one necessarily needs to take into account effects of the heliospheric interface. In this paper we propose a new model that is relatively simple but takes into account all major effects of the heliospheric interface. This model can be applied for analysis of backscattered La-alpha radiation data obtained on board of different spacecraft.
 \end{abstract}

\noindent\rule{8cm}{1pt}\\
{$^*$ $<$okat@iki.rssi.ru$>$}

\clearpage
\section*{Introduction}
\noindent
  The Solar system is moving through a partially ionized Local Interstellar Cloud (LIC) with relative velocity of $V_{LIC}=26.4$~km/s (Moebius~et~al. 2004). The main component of chemical composition of the LIC is atomic hydrogen (about $90 \%$). Interstellar atoms of hydrogen (H atoms) can deeply penetrate inside the heliosphere due to their large free path. These atoms effectively interact by charge exchange with protons of the interstellar cloud and the solar wind. Direct and indirect measurements of H atoms at the Earth orbit can provide substantial information on the properties of both the heliospheric interface and the Local Interstellar Medium (LISM). Theoretical modeling of distribution of the interstellar H atoms inside the heliosphere is fairly important to understand physical processes in the heliosphere and for correct interpretation of experimental data. At present main information on the properties of interstellar H atoms inside the heliosphere are obtained from backscattered Ly-alpha radiation measurements obtained by SOHO, HST, Voyager~1, Voyager~2, Pioneer-10, Cassini, Ulysses, Galileo and other spacecraft. The diagnostics of the H atom distribution by backscattered Lyman-alpha radiation is possible because spectral properties of the radiation considerably depend on distribution function of H atoms (Quemerais~2006).

   During several last decades the classical hot model is generally employed to analyze  backscattered Ly-alpha radiation data (Wu and Judge 1979; Lallement et~al.~1985). With this model one can obtain a solution of the Boltzmann's equation for the velocity distribution function of H atoms. However, the model has several crude assumptions that make the model less appropriate for correct analyzes of experimental data. These assumptions are:
   \begin{itemize}
     \item Maxwellian distribution function is taken at infinity (i.e. in LISM) as the outer boundary condition:
    \[
    \lim_{r\to\infty}f(\textbf{w}_{H},\textbf{r})=f_{M}(\textbf{w}_{H}).
    \]
     This condition means that no effects of the heliospheric interface can be taken into account in the model.
         The heliospheric interface effects can be taken only by artificial changes of the interstellar bulk velocity and temperature.
     \item The model is stationary. This means that no variations of the solar wind parameters with the solar cycle can be taken into account in the frame of the model.
     \item The model is axisymmetric. This means that the effects connected with the heliolatitudinal variations of the solar wind's parameters or interplanetary and interstellar magnetic field can be taken into account.
   \end{itemize}

   \begin{figure}[!t]
  \includegraphics[scale=0.8]{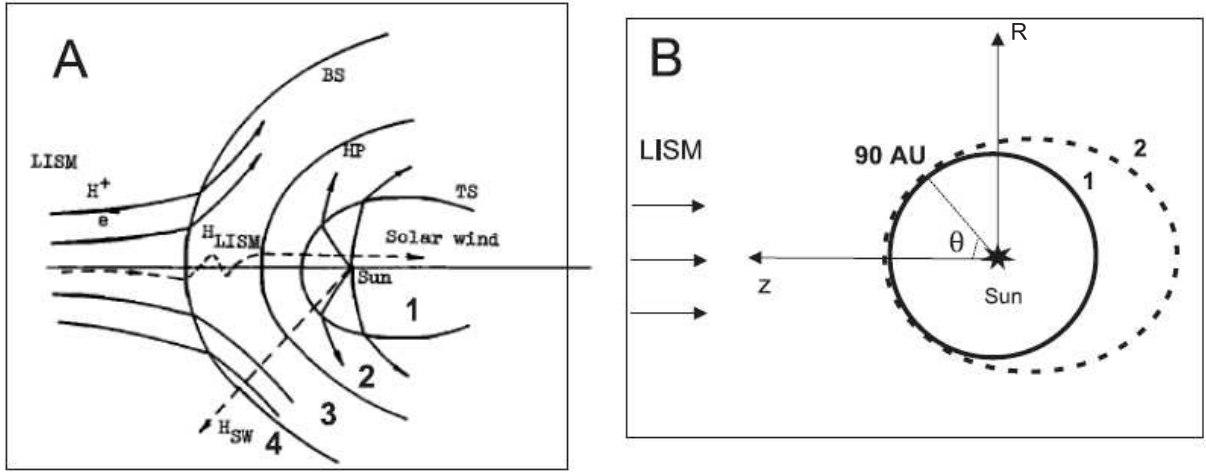}
  \caption{A: Qualitative structure of the heliospheric interface. HP is the tangential discontinuity (heliopause) separating the fully ionized solar wind from the interstellar plasma component, BS is the bow shock, and TS is the heliospheric termination shock. The dashed lines show the trajectory of a solar wind atom of populations 1 and 2 and of an interstellar atom of populations 3 and 4. B: cylindrical coordinate system that is employed in this work.
$z$-axis is the axis of symmetry, it is directed opposite to upwind; curve~1 is the outer boundary; curve~2 is the heliospheric termination shock. }
   \label{interface-sc}
\end{figure}
 In recent modifications of the hot model (Lallement et~al.~1991, Bzowski \& Rucinski~1995) the boundary conditions are taken far from the Sun (particularly at sphere with radius 80-90 AU), where solar gravitation, radiation pressure and photoionization are negligibly small.
Parameters (i.e. number density, average velocity and average temperature) of the Maxwellian velocity distribution function of H atoms at the boundary were used as free parameters and chosen in order to get optimal agreement between theoretical and experimental results. Costa~et~al. (1999) determined the temperature and velocity of interstellar atoms in the distant heliosphere (at 50 AU) on basis of analysis SWAN/SOHO data. It was shown that these values are significantly  different from local interstellar parameters known from the measurements of interstellar helium (Moebius et~al.~2004). These discrepancies are explained by heating and deceleration of hydrogen atoms's flow during their motion through the heliospheric interface (see Baranov and Malama~1993).
  Fig.~\ref{interface-sc}A demonstrates a qualitative picture of the heliospheric interface region. The heliopause is a contact discontinuity, which separates the solar wind and the interstellar plasma component. The termination
  shock (TS) is formed due to the deceleration of the supersonic solar wind. The bow shock (BS) may also exist if the interstellar plasma flow is supersonic. Four regions can be distinguished: the supersonic solar wind (region 1);
  the solar wind flow between the TS and the HP (region 2 or the inner heliosheath); the disturbed interstellar plasma component flow (region 3 or the outer heliosheath); the undisturbed interstellar gas flow (region 4).

   The first self-consistent kinetic-gasdynamic model of the solar wind interaction with the two-component (neutral atoms and plasma) LIC had been developed by Baranov and Malama (1993). In this model the kinetic equation for the distribution function of H atoms is solved self-consistently with the Euler equations for the charged component. The main physical process considered in the model is the charge exchange process of the H atoms with protons: $H^{+}$ + $H$ = $H$ + $H^{+}$. Atoms newly created by charge exchange have the velocity of their ion counterparts in charge exchange collisions. Therefore, the velocity distribution of these new atoms depends on the local plasma properties in the place of their origin. It is convenient to distinguish four different populations of atoms depending on the region in the heliospheric interface where atoms were formed. Population~4 are the primary interstellar atoms, which suffer substantial filtration in the heliospheric interface. The mean free path of the atoms with small velocities is smaller as compared with the fast atoms. Therefore, slow atoms are more processed by charge exchange (and ionization processes in general). This kinetic effect called the selection effect, results in asymmetry of the distribution function of population 4 at the TS (see Izmodenov et~al.~2001). Population~3 of H atoms in the heliosphere are the secondary interstellar atoms created in the outer heliosheath. Population~2 are the atoms created in the inner heliosheath between the heliopause and the termination shock. The number density of these atoms is significantly smaller as compared with the number densities of primary and secondary interstellar atoms. The population~2 has no major importance for interpretation of backscattered Lyman-alpha radiation experiments, but has major importance for the Interstellar Boundary Explorer (IBEX) spacecraft measurements that are specifically designed for measurements of this component. Population~1 consists of the atoms created in the region~1 of the supersonic solar wind. It can be noted, atoms of this population have velocity $\sim400$ km/s, that leads to a large doppler shift in line of Ly-alpha. Hence, atoms of population~1 do not add backscattered photons in Lyman-alpha.

     The mean free pathes with respect of charge exchange are comparable or larger than the characteristic size of the heliospheric interface for all introduced populations of H atoms (see, e.g., Izmodenov et~al.~2000). While the interstellar H atoms passing the interface before entering inside the supersonic solar wind region their distribution function is disturbed significantly due to the process of charge exchange. It was shown in Izmodenov et~al.~(2001), that distribution function of all populations is very different from Maxwellian at any given point of the heliospheric interface. Actually, the proof of the non-maxwellian behavior of the interstellar H atoms in the heliosphere was done by Baranov~et~al. (1998) who have shown that $T_{R}$ and $T_{z}$ temperatures are different. The temperatures are defined as:
   \[
 \begin{aligned}
 T_{R}(\textbf{r})&=\frac{m_{H}}{k\cdot n_{H}(\textbf{r})} \int f(\textbf{r},\textbf{w})\cdot (V_{R}(\textbf{r})-w_{R})^{2}d\textbf{w} \\
 T_{z}(\textbf{r})&=\frac{m_{H}}{k\cdot n_{H}(\textbf{r})}\int f(\textbf{r},\textbf{w})\cdot (V_{z}(\textbf{r})-w_{z})^{2}d\textbf{w}, \\
 \end{aligned}
\]
  here $f(\textbf{r},\textbf{w})$ is the velocity distribution function, $\textbf{r},\textbf{w}$ is the radius-vector of atom and its velocity, respectively. $n_{H}(\textbf{r})$ is the number density, $V_{R}$ and $V_{z}$ are components of the bulk velocity in the cylindrical system of coordinate (with $z$ axis as the axis of symmetry). $m_{H}$ is the mass of atomic hydrogen, $k$ is the Boltzmann constant. Significant difference between values of component $T_{R}$ and $T_{z}$ shown by Baranov et al. (1998) has proven that there is an asymmetry of the distribution function in different directions.

    In this work we explore the influence of the heliospheric interface on the distribution of the interstellar H atoms inside the heliosphere. To do this we compare the results obtained in the frame of the hot model with the results obtained in the frame of the heliospheric interface model (Baranov and Malama, 1993).
The comparison illuminate the role of the heliospheric interface on the distribution of H atom in the supersonic solar wind region.
   We also present an advanced hot model that combines the simplicity of the classical hot model with a possibility to take into account the heliospheric interface effects on the distribution of the H atoms.

\section*{Model}
\noindent
We study the velocity distribution function and its momentums of the interstellar H atoms inside the heliospheric termination shock (i.e. in the region 1 introduced above).
In the considered model we obtain the distribution function inside the sphere with center at the Sun and radius 90~AU. This sphere serves as the outer boundary. It is located inside the heliospheric termination shock and in upwind direction it is close to the TS (see Fig.~\ref{interface-sc}~B). We use the following notations: $ z $-axis is directed in the opposite direction of the interstellar flow, $R$-axis (in cylindrical coordinate system) is directed in the direction perpendicular to the $ z $-axis. $ \theta $ is the angle counted from positive direction of the $z$-axis. In this paper we restrict ourselves to the stationary axisymmetric problem ($z$ is a symmetry axis). This  corresponds to stationary and spherically symmetric flow of the solar wind. However, the method and numerical code that was developed by us can also be used for the non-stationary and three-dimensional cases without any changes.
\clearpage

  \begin{figure}[tp]
  \includegraphics[scale=0.8]{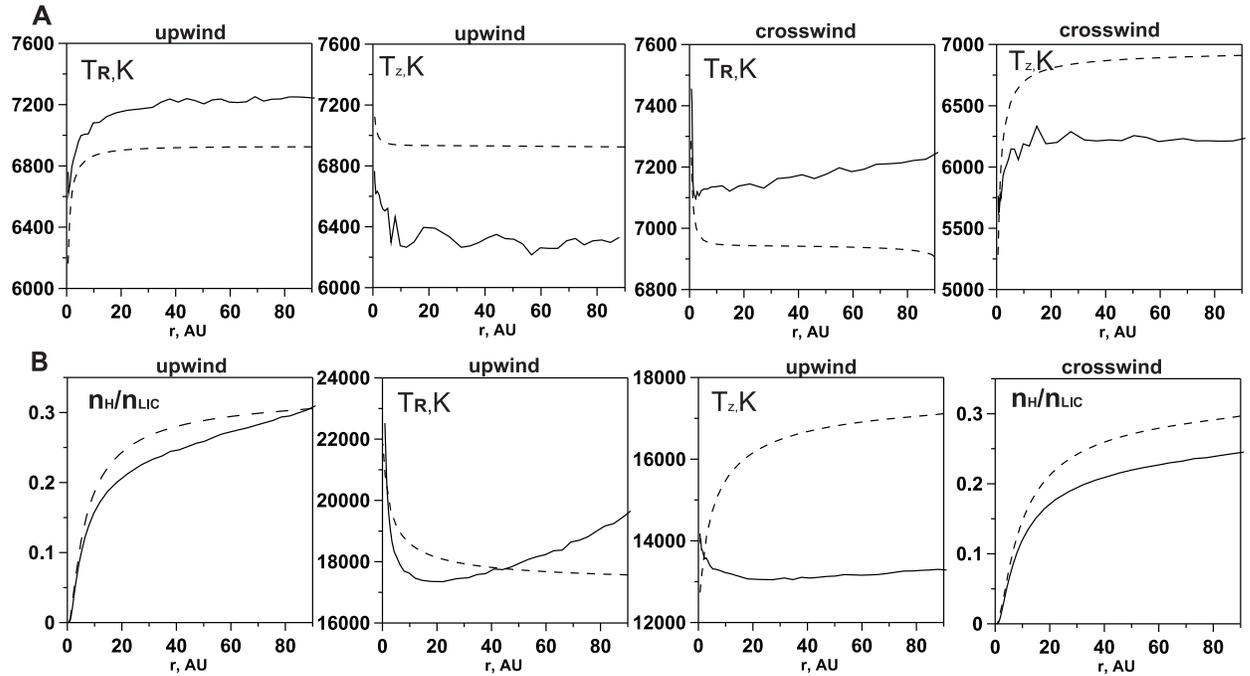}
  \caption{Comparison of the results of the hot and Baranov-Malama models. Top raw (A) shows parameters for the primary interstellar atoms: $T_{R}$ and $T_{z}$ temperatures in the upwind and crosswind directions. Bottom raw (B) shows parameters of the secondary interstellar atoms: number density (in dimensionless units) in upwind and crosswind, temperature $T_{R}$ in upwind and temperature $T_{z}$ in crosswind.
 Number density for primary H atom component is not presented because the curves of number density coincide for different models that makes the comparison not informative.
\newline
Solid curves correspond to the Baranov-Malama model; dashed curves correspond to hot model with Maxwellian distribution at the outer boundary.}
   \label{bm_cl}
\end{figure}

  In the solar system atom is affected by the solar gravitational force $\textbf{F}_{g}$ and the solar radiation pressure $\textbf{F}_{rad}$. These forces act in opposite direction and both are proportional to $1/r^{2}$. We can introduce dimensionless parameter
 \[
 \mu=|\textbf{F}_{rad}|/|\textbf{F}_{g}|,
\]
 which determines a balance between the solar gravitational and radiation pressure forces. Resulting force acting on unit of mass is represented as:
 \[
\textbf{F}=(1-\mu)\textbf{F}_{g}=-\frac{(1-\mu)GM_{s}}{r^{2}}\cdot \frac{\textbf{r}}{r},
 \]
here $G$ is the gravitational constant, $M_{s}$ is mass the Sun, $\textbf{r}$ is the radius-vector. In general case parameter $\mu$ depends on time and radial component of velocity $v_{r}$. However, for

 the purposes of this paper we assume that $\mu=1.258$ and this value does not change with time or atom velocity. Therefore, we will restrict the study to the case when the resulting force \textbf{F} is repulsive. The chosen value of $\mu$ is obtained by averaging of experimental data during one solar cycle.
The kinetic equation for the velocity distribution function of H atoms can be written as:
\begin{equation}
   \frac{\partial f(\textbf{r},\textbf{w})}{\partial t}
   + \textbf{w}\cdot\frac{\partial f(\textbf{r},\textbf{w})}{\partial \textbf{r}}
   + \textbf{F}\cdot \frac{\partial f(\textbf{r},\textbf{w})}{\partial \textbf{w}}
   + f(\textbf{r},\textbf{w})\cdot \frac{\partial \textbf{F}}{\partial \textbf{w}}
   = -\beta(r)\cdot f(\textbf{r},\textbf{w})\label{Bolcman}
\end{equation}
Here $\textbf{r},\textbf{w}$ are the radius-vector and the velocity of H atom, respectively; $f(\textbf{r},\textbf{w})$ is the velocity distribution function of H atoms; \textbf{F} is the force as described above. The last term in the left part of this equation appears in the 'exotic' case, when the effective force depends on the radial component of the H atom velocity (see Longmire~1963). For case when $\mu$ does not depend on the velocity of atom this last term is equal to zero. The right part of the kinetic equation includes the loss of atoms due to the charge exchange process, photoionization~($H+h\nu=H^{+}+e$) and electron impact ionization~($H+e=H^{+}+2e$).  Coefficient $\beta(r)$ is the effective ionization rate:
$\beta(r)=\beta_{ex}(r)+\beta_{ph}(r)+\beta_{ei}(r)$ , where $\beta_{ex}(r),\beta_{ph}(r) \mbox{ θ } \beta_{ei}(r)$ are the rates of the charge exchange, photoionization and electron impact ionization, respectively.
It is assumed in the model that the rates of charge exchange, photoionization and electron impact ionization decrease with distance from the Sun as $\sim1/r^{2}$, where $r$ is the heliocentric distance. This assumption deals with the fact that the rates
  $\beta_{ex}$, $\beta_{ei}$ and $\beta_{ph}$ are proportional to the number density of the solar protons, electrons and photons, respectively, and in the stationary case these number densities are proportional to $\sim1/r^{2}$. Thus:
  \[
 \beta(r)=\left( \beta_{ex,E}+\beta_{ph,E}+\beta_{ei,E} \right)\left(\frac{r_{E}}{r}\right)^{2} = \beta_{E}\left(\frac{r_{E}}{r}\right)^{2}.
  \]
Here $r_{e}=1$ AU is the distance from the Earth to the Sun. Symbol $E$ shows the values at the Earth's orbit. In this paper we assume $\beta_{E}=6.2\times 10^{-7}$ s$^{-1}$. This magnitude was obtained by averaging of available relevant experimental data over the solar cycle.

   Generally speaking, new hydrogen atoms (population~1) are created inside the computational domain due to the charge exchange process. These atoms have properties (in particulary the velocity and temperature) of the supersonic solar wind. Therefore, a photon scattered by such an atom has large doppler shift and does not contribute to the backscattered Ly-alpha spectrum. Thus, population~1 is not important for the present study and the charge exchange process results simply in losses of other populations of H atoms inside the heliosphere.

  The kinetic equation~\ref{Bolcman} is a linear partially differential equation that can be solved by the method of characteristic. The characteristics of equation~\ref{Bolcman} coincide with the ballistic trajectories of individual H atoms. The distribution function $f(\textbf{r}, \textbf{w})$ changes along a characteristic as following
 \[
 \frac{d {f} ( {\textbf{r}}, {\textbf{w}}, {t})}{d {t}} = -{\beta}({r})\cdot  {f} ( {\textbf{r}}, {\textbf{w}}).
 \]
After integration of the last equations taking into account the outer boundary conditions one can write the solution of equation~\ref{Bolcman} as follows:
 \begin{align}
   f( {\textbf{r}}, {\textbf{w}})=
  {f}_{S}( {\textbf{r}_{s}}, {\textbf{w}_{s}})\cdot
  \exp\left( -\int {\beta}( {r})d {t}\right)= \notag \\
  ={f}_{S}( {\textbf{r}_{s}}, {\textbf{w}_{s}})\cdot
  \exp\left( -\int \frac{ {\beta}_{E}}{{r}^{2}}d {t}\right).
 \end{align}
Here $f_{S}( \textbf{r}_{s}, \textbf{w}_{s})$ is the velocity distribution function of the H atoms at the outer boundary; $\textbf{r}_{s},\textbf{w}_{s}$ are the radius-vector and velocity vector at the intersection of the corresponding trajectory (characteristic) with the outer boundary sphere.
The integration is taken along the trajectory of atom (i.e. along the characteristic).
The hot model assumes the boundary conditions in form of the Maxwellian velocity distribution that can be written as:
 \[
f_{s}(\textbf{w})=  \frac{n_{H}}{\pi \sqrt{\pi} \cdot c_{s}^{3} }
 \cdot \exp \left( -\frac{\left(\textbf{V}-\textbf{w} \right)^{2}}{c_{s}^{2}} \right)  \mbox{,} \quad
c_{s}=\sqrt { \frac{2kT_{av}}{m_{H}} }.
\]
In the hot model case the number density ($n_{H}$), bulk velocity ($\textbf{V}$) and temperature ($T_{av}$) are constant at the boundary sphere.
In the calculations presented here the parameters were taken from the results of Baranov-Malama model at the point of crossing the 90 AU sphere and the z-axis (i.e. at the 90 AU sphere in the upwind direction).
Described above approach corresponds to the approaches of the models that were employed for the analysis of the backscattered Ly-alpha data. Such an approach takes into account of the heliospheric interface in the zero order of approximation. As it will be shown below such approach is not enough to get accurate results.
\section*{Comparison of the hot model and Baranov-Malama model results}
\noindent
   The comparison of the results obtained in the frame of the described above hot model with the results of the full self-consistent two component model by Baranov and Malama (1993) is performed in this section in order to estimate importance of the changes of the velocity distribution function in the heliospheric interface on the distribution of the H atoms in the region inside the termination shock. Fig.~\ref{bm_cl} shows parameters of primary (populations 4) and secondary (population 3) interstellar atoms as a function of the the heliocentric distance. The results are shown in the upwind and crosswind directions.
   As it can be seen from Fig.~\ref{bm_cl} that the Baranov-Malama model gives the significant differences between $T_{R}$ θ $T_{z}$ temperature components at 90~AU. This is the signature of the non-maxwellian behavior of the distribution function after passing the heliospheric interface.
    Since the standard hot model operates with the Maxwellian distribution at the boundary and therefore with the $T_{R}$ θ $T_{z}$ components equal each to other, the Figure~\ref{bm_cl} shows $5-10\%$ differences in $T_{R}$ and $T_{z}$ components obtained in the frame of the hot model and Baranov-Malama model for primary interstellar atoms. This temperature component difference remains from 90 AU upto the small heliocentric distances. For the secondary interstellar atoms (population~3) the distribution function is more different from Maxwellian as compared with the primary component. This leads to $10-20\%$ difference in the number densities of the hot and Baranov-Malama model for this component. Differences in the temperatures for this population are about $15\%$. Note, that there are qualitative difference in behavior of $T_{z}$ component. This fact is connected with nonzero third moments of distribution function of population~3 at the boundary sphere, which does not take into account in the hot model.

     Presented plots demonstrate that there are significant differences between the results of the standard hot model and Baranov-Malama model. This fact leads to the conclusion that the heliospheric interface considerably affects the distribution of the hydrogen atoms. The velocity distribution function and its moments in the supersonic solar wind are key factors to determine the spectrum and spectral properties of the backscattered Ly-alpha radiation inside the heliosphere. Therefore, application of the standard hot model for computation of spectrum of backscattered Ly-alpha radiation can lead to mistakes in analyzes of experimental data.

   Quemerais and Izmodenov~(2002) have calculated the spectrum of backscattered Ly-alpha radiation by employing the results of the Baranov-Malama model. Three populations (populations 2, 3 and 4) of H atoms where taken into account in this model. It was be shown in the paper that the secondary interstellar atoms (population~3), which are absent in the classical hot model, have significant influence on the Ly-alpha spectrum. Existence of populations 2 and 3  results in the asymmetry of the spectra with respect of its maximum. This emphasizes necessity to take into account multi-component non-Maxwellian nature of the velocity distribution function of the H atoms. In addition to that for more precise analysis of the experimental data it is necessary to employ more detailed model for the H atoms distribution. This model should take into account non-stationary and 3D effects connected with dependence solar wind parameters on time and heliolatitude. The time-dependent and heliolatitudinal effects were considered by Bzowski et~al.~(1995, 2008) and Bzowski, Rucinski~(1995). However, these models assumed Maxwellian distribution function at the outer boundary that is inaccurate. In the next section new model of the interstellar H atom transport in the supersonic solar wind is presented. This model combines the classical hot model approach with the results of the self-consistent Baranov-Malama model. Therefore, in the frame of this model it becomes possible to take into account both the effects of the heliospheric interface and the effects connected with temporal and latitudinal variations of the solar wind.

\section*{The advanced hot model of the hydrogen atoms motion into the heliospheric termination shock}
\noindent
\begin{figure}[!t]
  \includegraphics[scale=0.8]{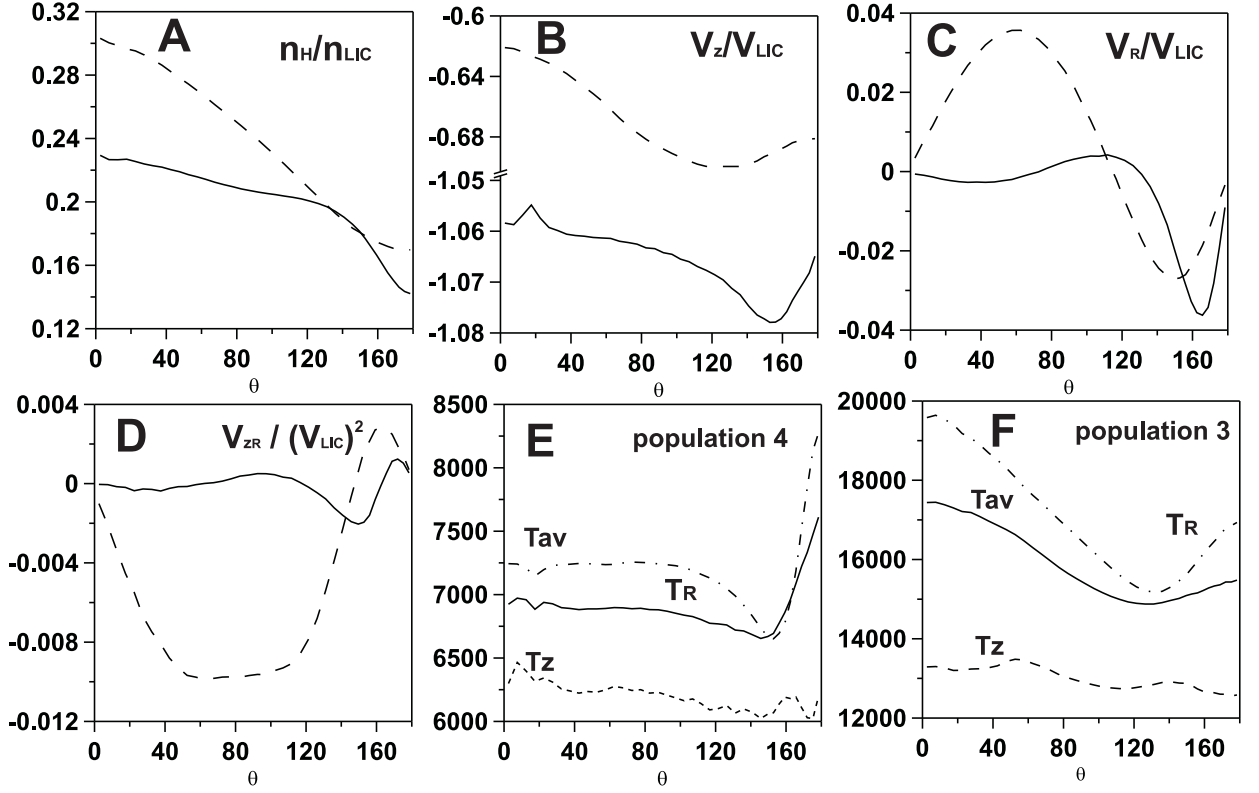}
  \caption{Moments of the velocity distribution function calculated in the frame of the  Baranov-Malama model as functions of the angle~($\theta$) that is measured from the upwind direction. All parameters except temperature are presented in dimensionless units. X-axis in these plots is  the angle $\theta$ in degrees.
Solid curves in plots A-D correspond to the primary interstellar atoms (population~4), dashed curves correspond to the secondary interstellar atoms of population~3. Plots E and F present temperatures: solid line - $T_{av}$, dash-dotted line - $T_{R}$ and dashed line - $T_{z}$ for the primary and secondary interstellar atoms, respectively.}
   \label{boundary}
\end{figure}
In order to take into account the effects of the heliospheric interface in the framework of the hot model, we use especial boundary condition for the velocity distribution function at outer sphere of 90 AU. This boundary condition takes into account Baranov-Malama model results. By doing this we simulate distribution functions of different populations of H atoms after they passed through the heliospheric interface region.

  From Baranov-Malama model we obtained values of zero, first and second moments of the distribution function. The moments are the number density $n_{H}$, bulk velocity $V_{R},V_{z}$, kinetic temperatures $T_{av}$, $T_{R}$, $T_{z}$ and the correlation coefficient $V_{zR}$ at outer boundary. Note, that these parameters depend on angle $\theta$ that is angle counted from the $z-$axis. They are also different for different populations of the H atoms. Correlation coefficient is defined as following:
    \[
       V_{zR}(\textbf{r})=\frac{1}{n_{H}(\textbf{r})} \cdot \int f(\textbf{r},\textbf{w})(V_{R}-w_{R})(V_{z}-w_{z})d\textbf{w}.
      \]
Here $\textbf{w}$ is the velocity of individual atom.
Dependence of the number density, velocity and temperatures of the H atom population on the angle $\theta$ is demonstrated in Fig.~\ref{boundary}. Note, that by definition: $3T_{av}=T_{R}+T_{z}+T_{\varphi},$ where $T_{R}, T_{z}, T_{\varphi}$ are the temperature components in the cylindrical coordinate system.
 From the axial symmetry of the considered problem we have $V_{\varphi}=0$ and as well as correlation coefficients $V_{R\varphi}=V_{z\varphi}=0$.
 In this work to get the H atom parameters at the outer boundary we run the Baranov-Malama model with the following boundary conditions in LISM: number densities of protons and neutral hydrogen atoms are $n_{p,LIC}=0.06 \mbox{cm}^{-3}, n_{H,LIC}=0.18~\mbox{cm}^{-3}$, respectively, relative velocity of the LISM is $V_{LIC}=26.4~\mbox{κμ/ρ}$, and temperature is $T_{LIC}=6519~K$. For the solar wind parameters at the Earth's orbit we assume the following numbers: $n_{p,E}=6~\mbox{cm}^{-3}, V_{E}=441.9~\mbox{km/s}$, and Mach number is assumed as $M_{E}=4.034$.

\subsection*{Analytical formulas for boundary condition}
\noindent
Different analytical formulas for the velocity distribution function with the given moments can be employed. In our calculations we used three different cases:
\begin{enumerate}
\item  Local-Maxwellian distribution function:
\[
f_{s}(\textbf{w})=  \frac{n_{H}}{\pi \sqrt{\pi} \cdot c_{s}^{3} }
 \cdot \exp \left( -\frac{\left(\textbf{V}-\textbf{w} \right)^{2}}{c_{s}^{2}} \right)  \mbox{,} \quad
c_{s}=\sqrt { \frac{2kT_{av}}{m_{H}} }
\]
with $n_{H}$,$\textbf{V}$,$T_{av}$ depending on the angle $\theta$.
\item  Three-temperature Maxwellian distribution with analytical formula that takes into account three kinetic temperature components ($T_{R}$,$T_{\varphi}$,$T_{z}$)
 \[
   f_{s}( \textbf{r}, \textbf{w} )= \frac {n_{H}} {c_{R}c_{\varphi}c_{z} \cdot \pi\sqrt{\pi} } \cdot \exp \left( - \left( \frac{(V_{R}-w_{R})^{2}}{c_{R}^{2}} + \frac{(V_{\varphi}-w_{\varphi})^{2}}{c_{\varphi}^{2}} + \frac{(V_{z}-w_{z})^{2}}{c_{z}^{2}} \right) \right)
 \]
 \[
    c_{R}=\sqrt { \frac{2kT_{R}}{m_{H}} }, c_{\varphi}=\sqrt { \frac{2kT_{\varphi}}{m_{H}} }, c_{z}=\sqrt { \frac{2kT_{z}}{m_{H}} }
 \]
\item Analytical formula for three-dimensional normal distribution that employs all zero, first and second moments of the distribution function including the correlation coefficient ($V_{zR}$). In this case the expression for the distribution function can be written as following:
\begin{equation}
 \begin{split}
 f_{s}(\textbf{r},\textbf{w})&=\frac{n_{H}}{(2\pi)^{3/2}\cdot \sqrt{(D_{R}D_{z}-V_{zR}^{2})D_{\varphi}}}\cdot \exp \Bigl(- \frac{1}{2} \Bigl( \frac{D_{z}}{D_{R}D_{z}-V_{zR}^{2}}(V_{R}-w_{R})^{2}+ {}\\
 &+ \frac{1}{D_{\varphi}}(V_{\varphi}-w_{\varphi})^{2}+ \frac{D_{R}}{D_{R}D_{z}-V_{zR}^{2}}(V_{z}-w_{z})^{2}- {}\\
 &- 2 \cdot \frac{V_{zR}}{D_{R}D_{z}-V_{zR}^{2}}(V_{R}-w_{R})(V_{z}-w_{z}) \Bigr)\Bigr)
 \end{split} \label{formula}
\end{equation}
Here $ D_{R}=\frac{k}{m_{H}}T_{R}, D_{\varphi}=\frac{k}{m_{H}}T_{\varphi},  D_{z}=\frac{k}{m_{H}}T_{z} $ and $k$ is the Boltzmann constant, $m_{H}$ is the mass of one hydrogen atom.
\end{enumerate}
In the latter case the outer boundary condition takes into account all second moments of the velocity distribution function, i.e. this case provides the most complete information on the velocity distribution function in comparison with other cases. In the rest of this section we will call model 1, model 2 and model 3 is accordance with the boundary condition case 1, 2 and 3, respectively.
\begin{figure}[!th]
  \includegraphics[scale=0.8]{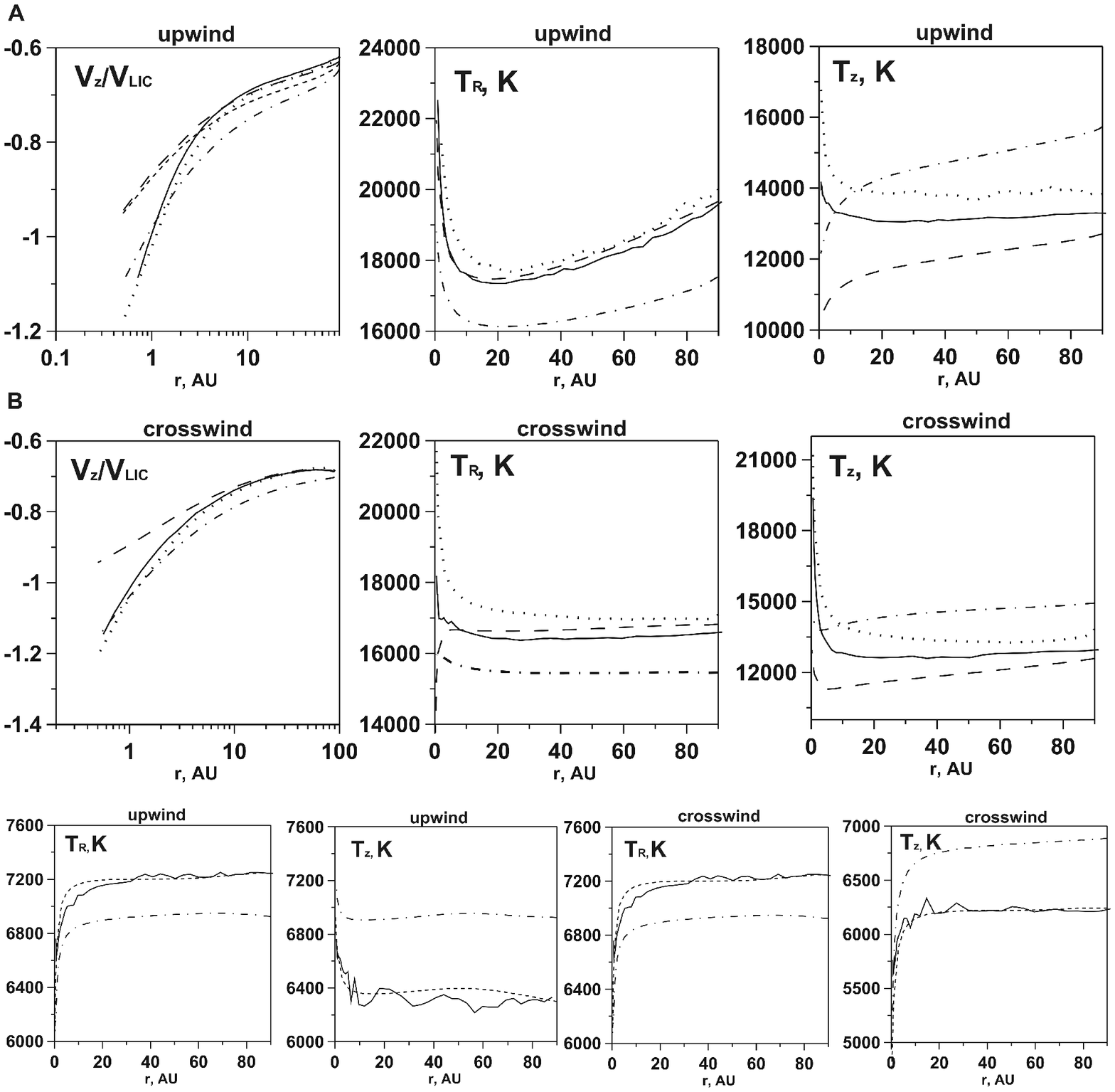}
  \caption{Comparison of the results of Baranov-Malama model with the advanced hot models. A: secondary interstellar atoms (population~3); B: primary interstellar atoms (population~4). Solid curves represent the results of the Baranov-Malama model, dashed-dotted curves show results for the model~1, dotted curves for the model~3. Small dotted curve in first plot in panel A shows results for model~2. Dots in panel A correspond to the results of model~4. Description of different models is given in the text.}
   \label{our_s3_s4}
\end{figure}
  Fig.~\ref{our_s3_s4} demonstrates the comparison of the results of the models 1, 2 and 3 with Baranov-Malama model results. As it can be seen from the plots, results of the simplest (first) model that takes into account dependence parameters at boundary sphere on angle $\theta$, but does not take into account non-Maxwellian nature of the distribution function, had the worst (among the three models) agreement with Baranov-Malama model's results. This fact emphasizes again that the distribution function of the primary and secondary interstellar atoms are substantially different from the Maxwellan even as far from the Sun as the termination shock, and these differences are noticeable in the distribution of atoms at small heliocentric distances.
  Note, that the results of model~2 and model~3 are practically coincide each with other. Therefore, results of model 2 are shown only for the distribution of the velocity $V_{z}$ for the secondary interstellar atom population. These results are shown because for $V_{z}$ there are about 5\% difference between results of the models 2 and 3. This difference can be explained by taking into account nonzero correlation coefficient in the third model, which especially affects the bulk velocity value. As it can be seen from all presented plots, there is a pretty good agreement between the results of Baranov-Malama model and the model~3.  However, despite on the good agreement for all other parameters the behavior of the temperature $T_{z}$ (for population~3) with the heliocentric distance in the upwind direction is qualitatively different in the case of Baranov-Malama model as compared with all hot type models 1, 2, and 3. This means, that even knowledge of all second moments of the distribution function is not enough to get good adequate description for the  non-Maxwellian nature of the distribution function for the secondary interstellar atoms of population~3. We suggest that this qualitative difference in the behavior of $T_{z}$ is mainly connected with the asymmetry of the distribution function at the outer boundary due to the selection effects that was discussed in Introduction. The realistic velocity distribution function of H atoms in the vicinity of the TS has nonzero third moments, but this fact is not taken into account in the considered above cases. In the next section  another step toward the most complete and realistic description of the velocity distribution function at outer boundary will be done.
\subsection*{On the difficulties by using the distribution function from Monte-Carlo calculations as the outer boundary conditions of the hot-type models}
\noindent
  For the most correct determination of the distribution function at the TS that takes into account all effects of the heliospheric interface we can in principle use the velocity distribution function calculated directly in the frame of the Baranov-Malama model. Indeed, the Monte-Carlo method with splitting of trajectories (see Malama~1991) that is employed for the Baranov-Malama model enables to calculate the velocity distribution function as it was demonstrated by  Izmodenov~et~al. (2001). However, in order to get the velocity distribution with high order of accuracy that is necessary for the boundary condition of the hot model requires a very large number of the the trajectories in the Monte-Carlo methods and it requires very expensive (in time and power) computer simulations. Also the limitations of computer operation memory becomes a strong restriction. That is why the Baranov-Malama model can not be directly used for the calculation of the detailed velocity distribution function in the entire heliosphere. Nevertheless,  for the calculations of the backscattered Ly-alpha radiation spectral properties it is necessary to know the distribution function of H atoms. The knowledge of the moments is not enough.

   In addition to the models 1-3 discussed above we will consider model~4 that employs the velocity distribution function calculated in the frame of the Baranov-Malama model by Monte-Carlo method. In this case the distribution function at outer boundary is obtained numerically and is provide in the nodes of mesh, i.e. we know values of the velocity distribution function for the given values of angle $\theta$ and for the given three components of the velocity $w_{r}, w_{\theta}, w_{\varphi}$. The components are in the spherical coordinate system. Using of a grid of uniform distribution of the grid nodes in the velocity space leads to the insufficient results, because only the atoms with small angular component of velocity can reach vicinity of the Sun. Therefore, small numerical errors in the interpolation in the velocity space at the outer boundary may lead to the essential errors near to the Sun. This explains why we choose a nonuniform mesh in the velocity space. This grid has smaller cells when the angular components of the atom velocity are close to zero. Results obtained in the frame of the model~4 are presented on Fig.~\ref{our_s3_s4}. It can be seen that for model~4 the T$_z$ behaves qualitatively similar with the Baranov-Malama model. Let us remind that for the models 1-3 we had qualitative difference in the distribution $T_{z}$ for the secondary interstellar atoms as compared to the Baranov-Malama model. Despite that $T_z$ behaves better for model~4, all results obtained in the frame of model~4 are somewhat different for Baranov-Malama results. These are due to the quantitative errors connected with insufficient accuracy in the calculations of the distribution function by Monte-Carlo method, and \emph{essentially} with the inaccuracies in the interpolation of the distribution function at the outer boundary. Note that for the primary interstellar atoms (population~4) the results of model~3 have better agreement with Baranov-Malama model than for the model~4. This is connected with the fact that the distribution function of the primary interstellar atoms far from the Sun is quite close to the three-temperature Maxwellian distribution function. Also model~3 with analytical boundary conditions has no problems with interpolation errors, which appears in model~4 with tabular distribution function at the outer boundary.

     In conclusion of this section we summarize that:
  1)~use of the third advanced hot model (model~3) with analytical formula~(\ref{formula}) for the boundary conditions, which takes into account all second moments of distribution function, leads to acceptable agreement with the Baranov-Malama model;
  2)~for more precise accounting the global effects of the heliospheric interface it is possible to use realistic distribution function calculated by Monte-Carlo in the frame of Baranov-Malama model.
  However, in this case there are problems with accuracy of calculations. The problems are connected with the interpolation of the distribution function at the outer boundary
  between mesh's nodes in the angle $\theta$ and in the three components of velocity and also due to statistical errors of the Monte-Carlo method.
  All these facts do not allow to achieve good agreement of model~4 results with the results of the Baranov-Malama
  model despite the pretty good qualitative agreement of the model results.

 \section*{Conclusions}
\noindent
In this work we presented the comparison of the results obtained in the frame of the standard hot model that does not take into account effects of the heliospheric interface with the results of the kinetic-gasdynamic Baranov-Malama model. Some possibilities for advancement of the hot model are presented and discussed. In order to improve the hot-type model we need to employ the specific boundary conditions at the outer boundary far from the Sun. These  boundary conditions should take into account the changes of the distribution function of hydrogen atoms in the heliospheric interface region. Different cases of the analytical boundary condition were considered. Our main conclusions are following:
  \begin{enumerate}
     \item The modifications of the velocity distribution function of the H atoms in the heliospheric interface region has essential
     influence on the distribution of the interstellar H atoms in the heliosphere. These heliospheric interface effects leads to the substantial differences between the results of the hot and kinetic-gasdynamic Baranov-Malama models.
     \item It was shown in the paper that the advancing the hot model by using local Maxwellian distribution function at the termination shock (i.e. Maxwellian distribution with parameters, which depends on angle $\theta$) is not enough for correct inclusion of the heliospheric interface effects.
     \item It was shown that model~3 with the outer boundary condition (\ref{formula}), which takes into account all zero, first and second moments of the velocity distribution at the outer boundary, provides the best agreement with the results of the Baranov-Malama model.
       Main difference between the results of the model~3 and Baranov-Malama model is in the qualitative difference in the temperature $T_{z}$ component of the secondary interstellar atoms in the upwind direction. This means than knowledge of all zero, first and secondary moments of the distribution function at the termination shock is not enough to obtain complete adequate description of non-maxwellian character of the distribution function of the secondary interstellar atoms of population~3.
     \item Using velocity distribution function obtained numerically in the framework of Baranov-Malama model does not allow to get good agreement because of computational restrictions connected with the precisions of the Monte-Carlo method and accuracy of the interpolation procedure. However, unlike the models 1-3 the model~4 provides qualitative agreement of $T_{z}$ component distribution for the secondary interstellar atoms with the results of the  Baranov-Malama model.
    \end{enumerate}

\pagebreak

\clearpage
\pagebreak


\end{document}